\documentclass[11pt]{article}

\usepackage{graphicx,epsfig}
\usepackage{multicol}
\usepackage{amsmath,amssymb,cite,color,hyperref}
\usepackage{lineno}

\newcommand{\be}{\begin{equation}}
\newcommand{\ee}{\end{equation}}
\newcommand{\bea}{\begin{eqnarray}}
\newcommand{\eea}{\end{eqnarray}}
\newcommand{\bi}{\begin{itemize}}
\newcommand{\ei}{\end{itemize}}
\newcommand{\ben}{\begin{enumerate}}
\newcommand{\een}{\end{enumerate}}

\definecolor{grey}{rgb}{0.5,0.5,0.5}

\def\iab{ab$^{-1}$}
\def\ifb{fb$^{-1}$}
\def\ipb{pb$^{-1}$}
\def\lum{cm$^{-2}$s$^{-1}$}
\def\tev  {\ifmmode {\mathrm{TeV}} \else TeV\fi}

\begin{document}

\title{\textbf{\sc Luminosity goals for a 100-TeV pp collider}}

\author{
Ian Hinchliffe$^a$\footnote{\href{mailto:i_hinchliffe@lbl.gov}{i\_hinchliffe@lbl.gov}}, Ashutosh Kotwal$^{b}$\footnote{\href{mailto:kotwal@phy.duke.edu}{kotwal@phy.duke.edu}}, Michelangelo
L. Mangano$^c$\footnote{\href{mailto:michelangelo.mangano@cern.ch}{michelangelo.mangano@cern.ch}}, Chris Quigg$^d$\footnote{\href{mailto:quigg@fnal.gov}{quigg@fnal.gov}}, Lian-Tao Wang$^e$\footnote{\href{mailto:liantaow@uchicago.edu}{liantaow@uchicago.edu}} \\[5pt]
~\\
\small
$^a$ Phyiscs Division, Lawrence Berkeley National Laboratory, Berkeley
  CA 94720, USA\\[5pt] 
\small
$^b$ Fermi National Accelerator Laboratory, Batavia, Illinois 60510, USA
\\ 
\small
Duke University, Durham, North Carolina 27708, USA\\[5pt]
\small
$^c$ PH Department, TH Unit, CERN, CH-1211 Geneva 23, Switzerland\\[5pt]
\small
$^d$ Theoretical Physics Department, Fermi National Accelerator
 Laboratory\\ 
\small
P.O. Box 500, Batavia, Illinois 60510 USA\\ 
\small
Institut de
 Physique Th\'eorique Philippe Meyer, \'Ecole Normale Sup\'erieure\\ 
\small
24 rue Lhomond, 75231 Paris Cedex 05,
 France\\[5pt]
\small
$^e$ Department of Physics and Enrico Fermi Institute, 
University of Chicago, Chicago, IL 60637
USA}


\maketitle
\date{}

\vspace{1cm}
\begin{abstract}
We consider diverse examples of science goals that provide a framework to assess luminosity goals for a future 100-TeV
proton-proton collider. 
\end{abstract}
\vfill
\begin{flushleft}
CERN-PH-TH/2015-089 \\ FERMILAB-CONF-15-125-E-T \\LBNL-176221 \\
\end{flushleft}

\section{Introduction}
\label{sec:intro}
All experimental measurements benefit from larger data sets, since
statistical uncertainties diminish. Some
measurements are ultimately limited by backgrounds or by systematic
uncertainties, but additional data can help to reduce these, or provide
alternative and independent measurements. For example, multipurpose
experiments such as those carried out at particle colliders (from the
$B$ factories up to the highest energy hadron colliders) explore a
broad spectrum of available observables, including those that have
very small rates that will continue to benefit as data sets increase.

Nevertheless, practical, technical, and financial considerations limit
the integrated luminosity that an accelerator will ultimately be able
to deliver, so it is important both to aim high and to anticipate what
the minimum luminosity must be to guarantee significant new
results. The measurement of specific processes can be used to define
such minimal goals. This is a well-posed problem in the case of
measurements of known processes, where the goal is, for example, a
given precision. In the case of searches for new phenomena, things are
less clear.  The searches for the top quark and the Higgs boson, whose
mass ranges and properties were well defined, set reliable luminosity
requirements that were used in setting the accelerator specifications
of the Tevatron, of its Run 2 upgrade, and of the Large Hadron
Collider. But after the Higgs discovery, we lack a well-defined
direction for the appearance of new physics phenomena that can be
guaranteed (or at least anticipated with a high degree of
confidence). Discoveries in Run 2 of the LHC and beyond could change
this situation.

The absence of a clear target leads, for now, to large uncertainties
in the definition of discovery-driven parameters of future colliders.
This is true both of possible discoveries at the highest mass reach
and of discoveries that might result if deviations from the standard
model were seen in precision studies of electroweak observables, or of
Higgs decays. In both cases one should {simply} aim at the most
aggressive possible performance (in energy and luminosity) allowed by
the balance of technological challenge and costs and then assess the
impact of such measurements. The impact must be large enough to both
motivate the experimental community to participate and justify the
cost of undertaking a major new project.

As the high energy physics community starts discussing scenarios for
future hadron colliders in the energy range of
$100~\tev$~\cite{benedikt,SPPC}, it is
natural to ask what the appropriate luminosity goals should be. A
generic argument, based on the scaling properties of cross sections as
a function of the partonic center-of-mass energy suggests that in
order for the increase in discovery reach to match the increase in
collider energy, $\sqrt{s}$, the luminosity should scale as $s$, the
square of the center of mass
energy~\cite{Barletta:2014vea,Richter:2014pga}. Scaling violations in
the partonic densities can be used to support an argument for even
faster luminosity growth~\cite{Mangano-IHEP,Rizzo:2015yha}. This
scaling argument has the virtue of simplicity, but the conclusions are
sensitive to the choice of starting parameters. It is worth recalling
that, because of the fixed size of the LEP tunnel, the LHC compensated
for constrained energy by setting aggressive luminosity goals. In
different circumstances, the energy--luminosity optimization might
take a different path.

In this note, we consider from a broader perspective the physics
opportunities that a 100-TeV hadron collider should address, among
them, extending the mass reach for discovery. Specifically, we examine
several physics cases that drive the luminosity goals. In the context
set by those goals, we ask how high a luminosity is desirable and
whether we can reasonably set a minimum acceptable
luminosity~\cite{Eichten:1984eu}. 

We set as a first requirement that \textit{the initial luminosity of a
  new hadron collider should be sufficiently high to surpass the
  exploration potential of the LHC very quickly,} certainly within the
first year of operation. We consider the luminosity demands of four
areas of investigation.
\begin{enumerate}
\setlength{\itemsep}{-4pt}
\item The search for new phenomena, inaccessible to the LHC, at high mass scales;
\item Increased sensitivity to rare or high-background processes at mass scales
  well below the kinematical limit of the 100 TeV collider;
\item Increased precision for studies of new particles within the ultimate
  discovery reach of the LHC;
  \item Incisive studies of the Higgs boson, both in the domain of
  precision, and in the exploration of new phenomena.
\end{enumerate}
\section{Luminosity Needs of the Physics Criteria}
\subsection{Extending the discovery reach at high mass scales}
\begin{figure}[htb]
   \centering
\epsfig{height=0.37\textwidth,figure=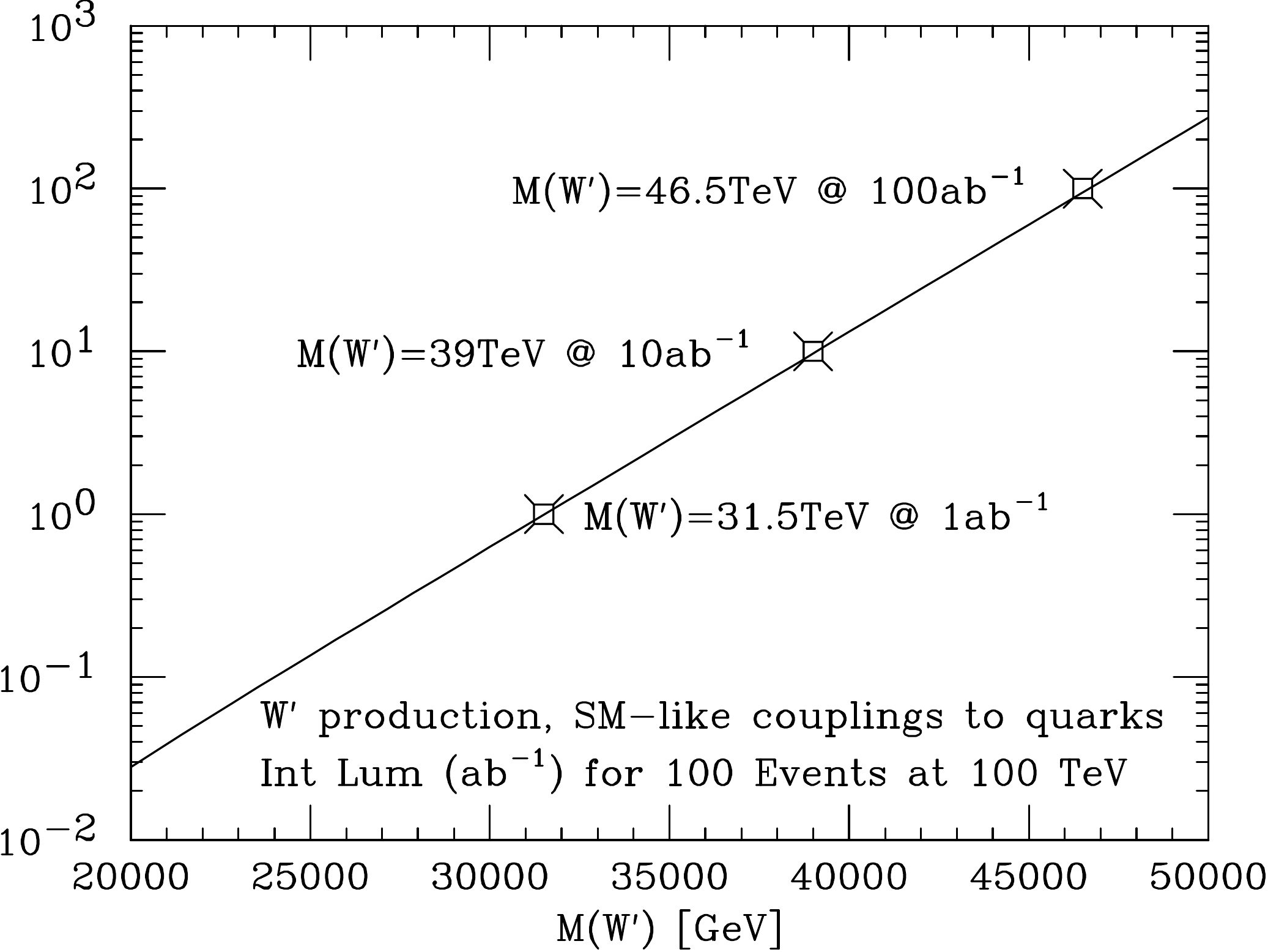}
\hfill
\epsfig{height=0.37\textwidth,figure=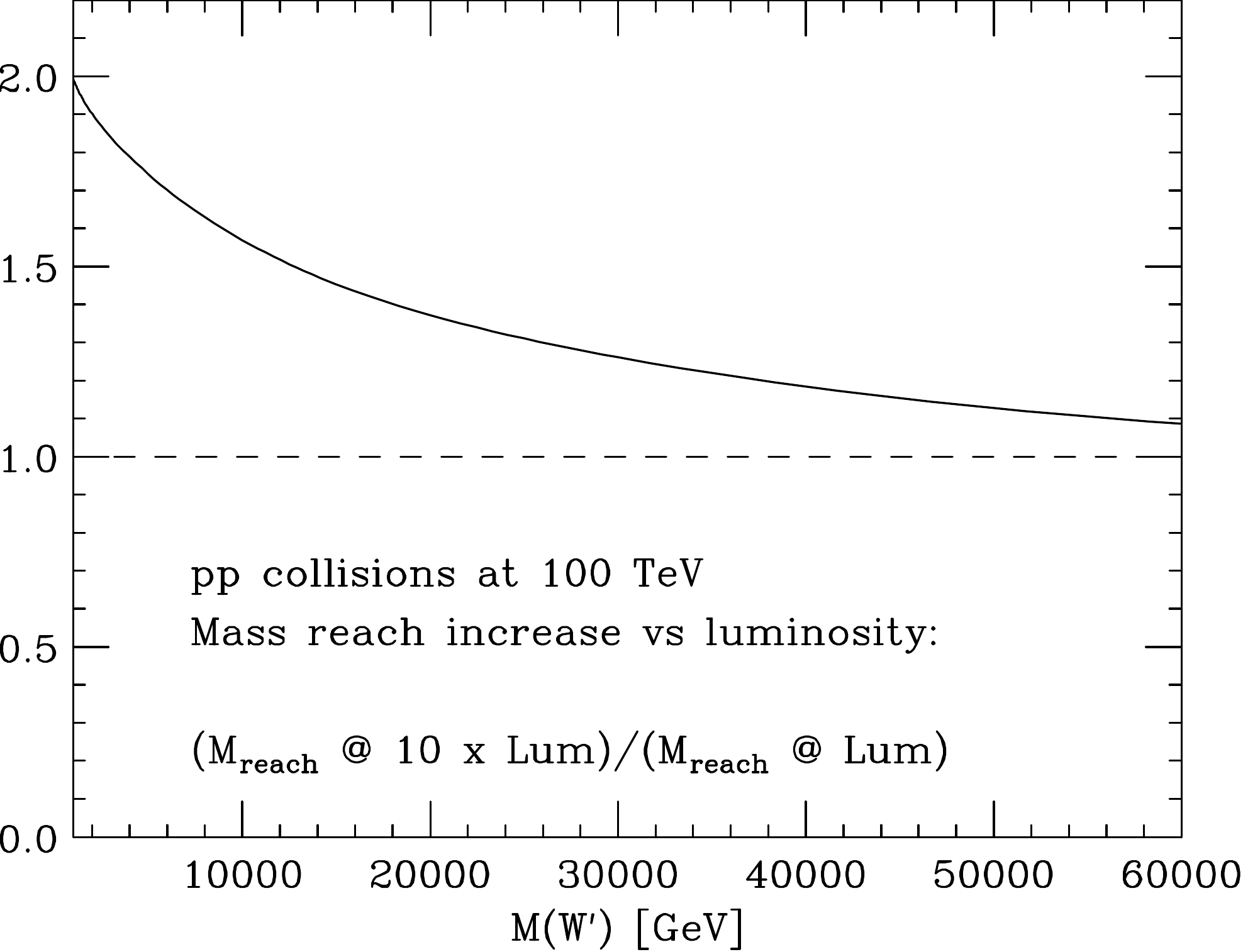}
   \caption{\small  Left plot: integrated luminosity (\iab) required to
     produce 100 events of a sequential standard-model $W^\prime$
     boson at 100 ~TeV, as a
     function of the $W^\prime$ mass. Right plot: mass reach increase for a
     sequential $W^\prime$ from a factor of 10 increase in luminosity.}
   \label{fig:Wreach}
\end{figure}

\begin{figure}[htb]
   \centering
\epsfig{height=0.37\textwidth,figure=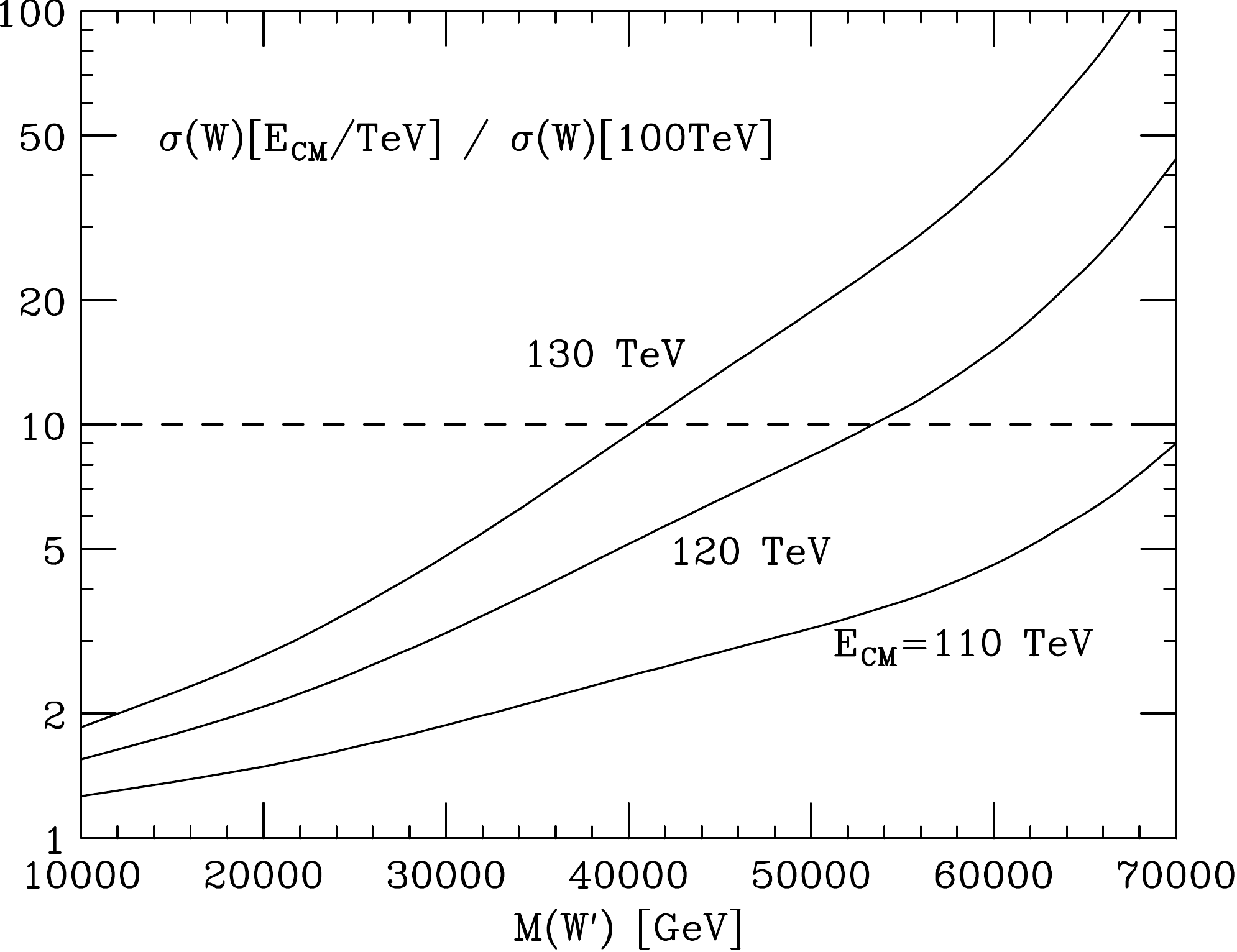}
\hfill
   \caption{\small  Ratio of $W^\prime$ production cross sections at
     different values of $\sqrt{s}$ to those at $\sqrt{s}=100$~TeV, as a function of the  $W^\prime$ mass.}
   \label{fig:Wenergy}
\end{figure}

We consider, as a first example, the case of a possible sequential
$W^\prime$ boson,
 a massive electroweak gauge boson with couplings identical to those of the
standard-model $W^\pm$ boson.  The production proceeds via quark anti-quark annihilation ($q\bar{q}$). Setting the discovery threshold at 100 total produced $W^\prime$
bosons (leading to $\sim 20$ events in the clean and background-free
leptonic final states with electrons and muons) gives the
luminosity requirements displayed in the left plot of Fig.~\ref{fig:Wreach}, as a
function of the $W^\prime$ mass $M(W^\prime)$. \footnote{The $W'$ cross
sections are calculated at LO, using the PDF sets
CTEQ6.6~\cite{Nadolsky:2008zw}  and scale
$Q=M_{W'}$.} 
In the luminosity range of
0.1--10$^3$~\iab, the increase in mass reach is well approximated by a
logarithmic behaviour, with a $\sim$ 7~\tev\ increase in mass for a tenfold
luminosity increase: $M(L)-M(L_0) \sim~7~\tev \, \log_{10}(L/L_0)$ (a
simple argument for this scaling relation is given in Appendix A).  
The
relative gain in mass reach therefore diminishes as the total
luminosity is increased, as shown in the right plot of
Fig.~\ref{fig:Wreach}. This displays the relative extension in mass reach
achieved with a factor of 10 increase in luminosity. For example, if
for a given integrated luminosity $L_0$ we are sensitive to a mass
$M_{W'}=20$~TeV, $10\times L_0$ will give sensitivity to a mass a
factor of $\sim 1.4$ times larger, namely 28~TeV.
The additional sensitivity gain given by a factor of 10 increase in
luminosity drops below 20\% at around 40~\tev,
the discovery reach corresponding to about 10~\iab\ (see the left plot
of Fig.~\ref{fig:Wreach}).
The conclusion  is that higher luminosity is of greater benefit
in the exploration of lower, rather than higher, masses.
To illustrate the interplay between collider energy and luminosity, we show in 
Fig.~\ref{fig:Wenergy} how cross sections increase as the c.m.\ energy is raised above $\sqrt{s}=100$~TeV. For a mass of
40~TeV, an increase in energy from 100~TeV to 130~TeV would be equivalent to a factor of 10 increase in luminosity at $\sqrt{s}=100$~TeV. 

\begin{figure}[htb]
   \centering
\epsfig{height=0.37\textwidth,figure=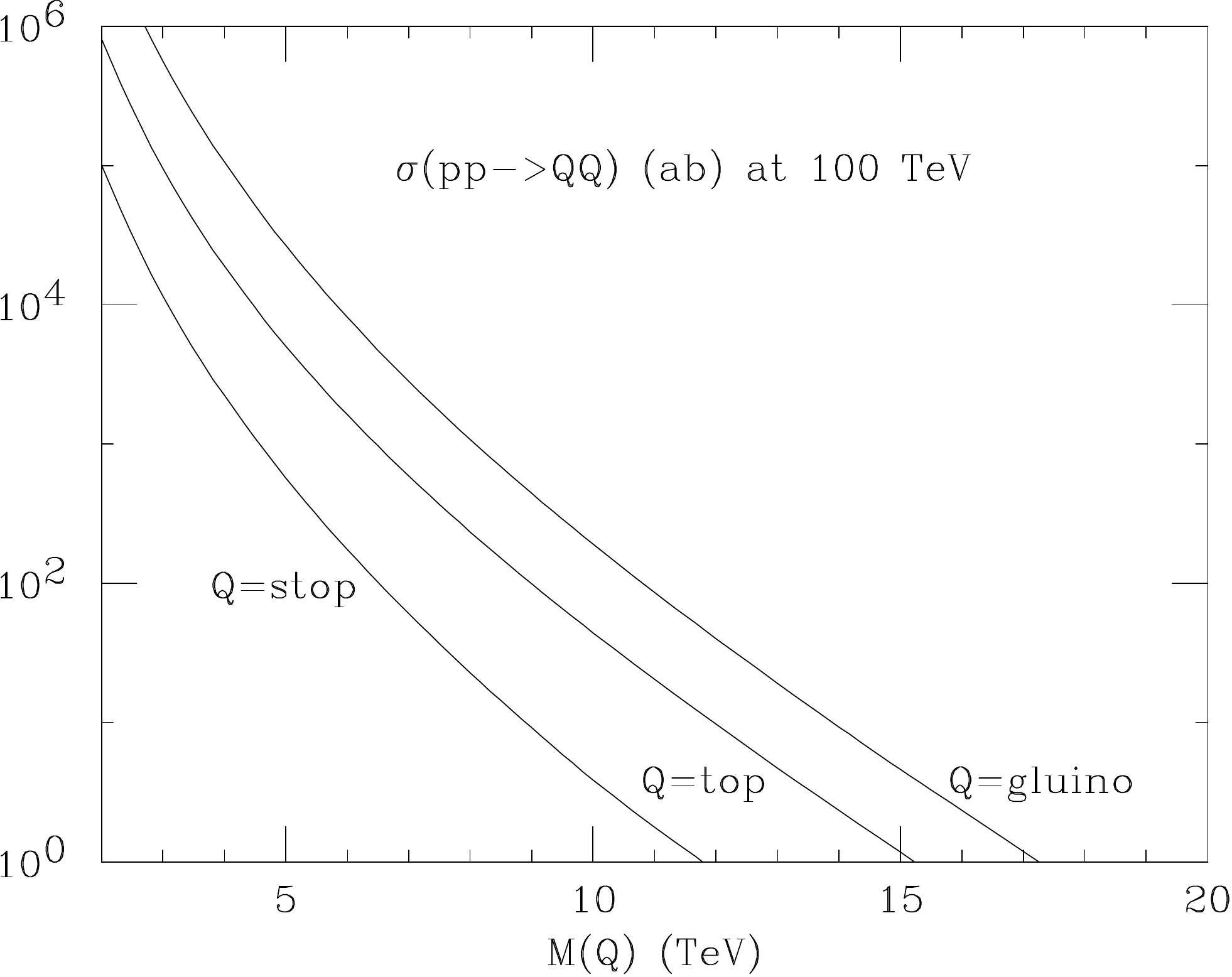}
\hfill
\epsfig{height=0.37\textwidth,figure=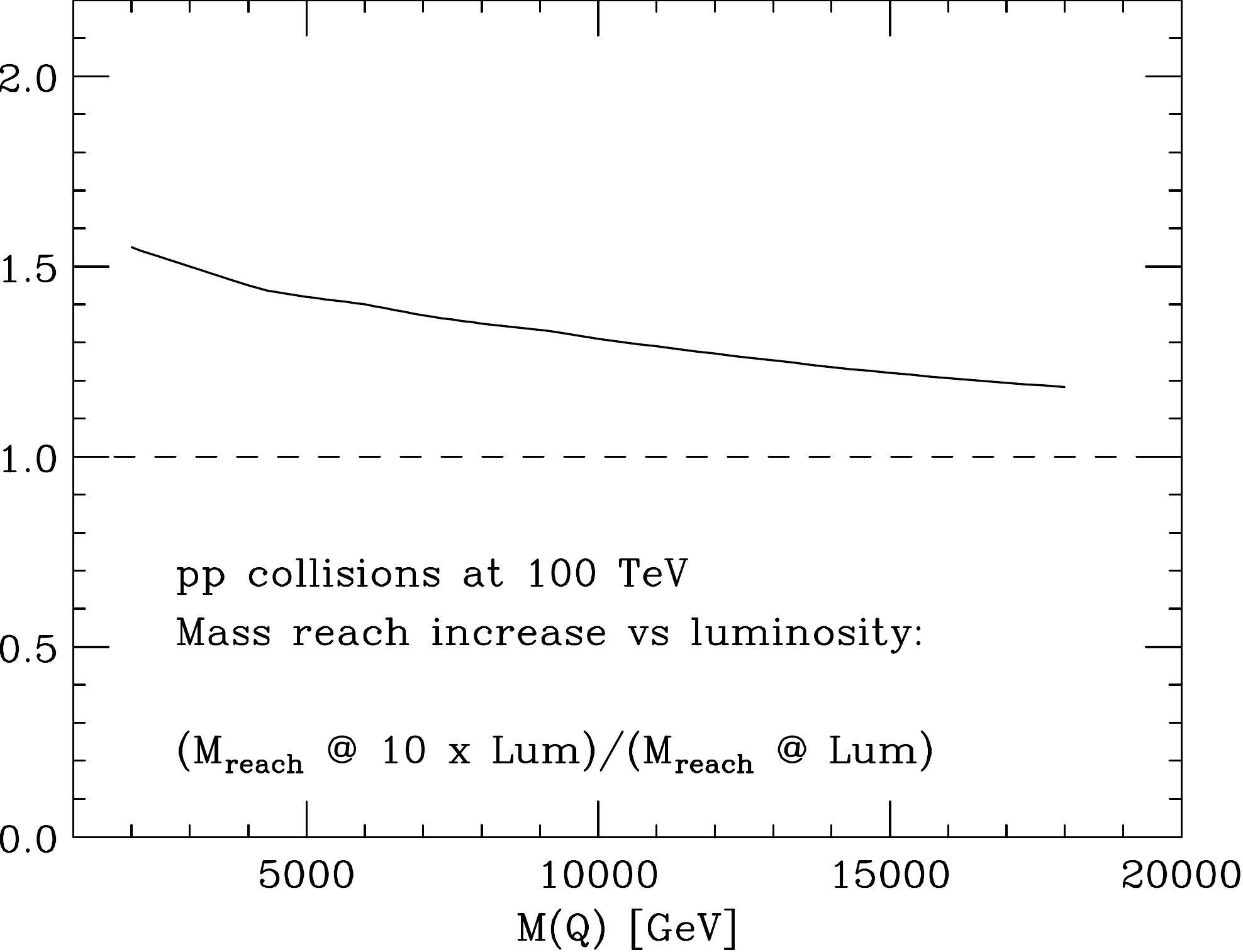}
   \caption{\small Left plot: cross sections for
     pair-production of colour-triplet scalars (``stop''), fermions (``top'')
     and gluinos, as a function of their mass.
     Right plot: mass reach increase for heavy
     quark pair production from a factor of 10 increase in luminosity. }
   \label{fig:QQreach}
\end{figure}

Qualitatively similar conclusions can be reached considering processes
dominated by a $gg$ initial state, rather than $q \bar{q}$. The
pair-production of massive color-triplet quarks and
squarks, and of gluino-like states, is shown in Fig.~\ref{fig:QQreach}.
As exhaustive list of additional examples is given in Ref.~\cite{Rizzo:2015yha}. 
\begin{table}[h]
\centering
 \begin{tabular}{|l|c|c|}
 \hline
 Integrated Luminosity & 300~\ifb & 3000~\ifb \\
 \hline
95\% CL exclusion limit (ATLAS~\cite{ATLAS:2013hta}) & 6.5~\tev & 7.8~\tev \\
5$\sigma$ discovery limit (CMS~\cite{CMS:2013xfa}) & 5.1~\tev & 6.2~\tev \\
 \hline
 \end{tabular}
\caption{\small 
Projected sensitivity, at $\sqrt{s} = 14$~\tev, for the exclusion and
discovery of a $Z^\prime$ gauge
boson with standard-model couplings. 
\label{tab:Zprime} }
\end{table}

The above qualitative analysis can be illustrated using more complete
studies done for the LHC luminosity upgrade, as shown for example in
Table~\ref{tab:Zprime}, which gives ATLAS and CMS's estimates for the
exclusion and discovery reach of a sequential standard-model
$Z^\prime$ gauge boson decaying to leptons. The mass reach increases
by 20\% as the integrated luminosity increases from 300 to
3000~\ifb. One could therefore argue that, from the perspective of
simply increasing the mass reach at the high end, the LHC will already
have almost saturated its discovery potential after 300~\ifb. Indeed
the main motivations for its upgrade to 3000~\ifb\ come from the need
to study with greater statistics the Higgs boson, or to search in
greater detail for elusive signatures of beyond-the-standard-model
phenomena in the TeV mass region (see, e.g., the studies performed in
the context of the ECFA Workshop on HL-LHC~\cite{ECFA-HL-LHC}).
Assuming 300~\ifb\ as a reference to scale the luminosity by the
factor of $s$, we obtain a target integrated luminosity of $300 \times
(100/14)^2$~\ifb $\sim 15$~\iab, a figure consistent with the current
parameters of the FCC-hh machine design~\cite{benedikt}.


\subsection{Enhancing the discovery reach at low mass}
By low mass we mean masses, or parton subenergies $\sqrt{\hat{s}}$, small relative to the kinematical
limit of the collider, $\sqrt{s}$: this category would include  the
top quark and the Higgs boson, as well as new particles such as
sleptons or charginos. For these particles the discovery can be limited by
the smalless of the cross sections, by the rarity of, or low efficiency for the
signal, by large backgrounds, or by important systematic uncertainties. The
discussion of the optimal luminosity is therefore very much
dependent on the process and on what the limiting factors are in its case.

If backgrounds are negligible, the searches for rare or
forbidden decays of a given particle, or for new particles with
low-rate but clean signatures, will benefit linearly from an increase
in luminosity.\footnote{Examples could include pair production of
  doubly-charged Higgses, decaying to final states like $e^+e^+
  \mu^-\mu^- + X$, or FCNC top decays such as $t\to c H$, with
$H\to\gamma\gamma$ or $\mu^+\mu^-$.} The required
amount of luminosity depends on the specific rate targets
that make these specific processes interesting. No general statement
can be made, and arguments such as scaling the luminosity
proportionally to $s$
do not necessarily apply.

How the discovery reach improves for low-efficiency and
large-background final states, {\it e.g.,} searches that rely on small
missing-$E_T$ signatures, is strongly affected by the detector
performance. Improvements in sensitivity from increasing statistics
through higher instantaneous luminosity will be limited when
systematics uncertainties dominate. Clear examples appear in the
projections being made for the HL-LHC. For example,
Fig.~\ref{fig:sbottom} shows the discovery and exclusion reach for
bottom squarks at the LHC, at 300 and 3000~\ifb, using $\tilde{b}\to
b\chi_0$ decays. The reduced sensitivity to final states with small
missing $E_T$ strongly limits the possible progress in the regions
of parameter space corresponding to compressed mass spectra, which
are shown to the right of the "forbidden" line on this plot. 
\begin{figure}[htb]
   \centering
\epsfig{width=0.6\textwidth,figure=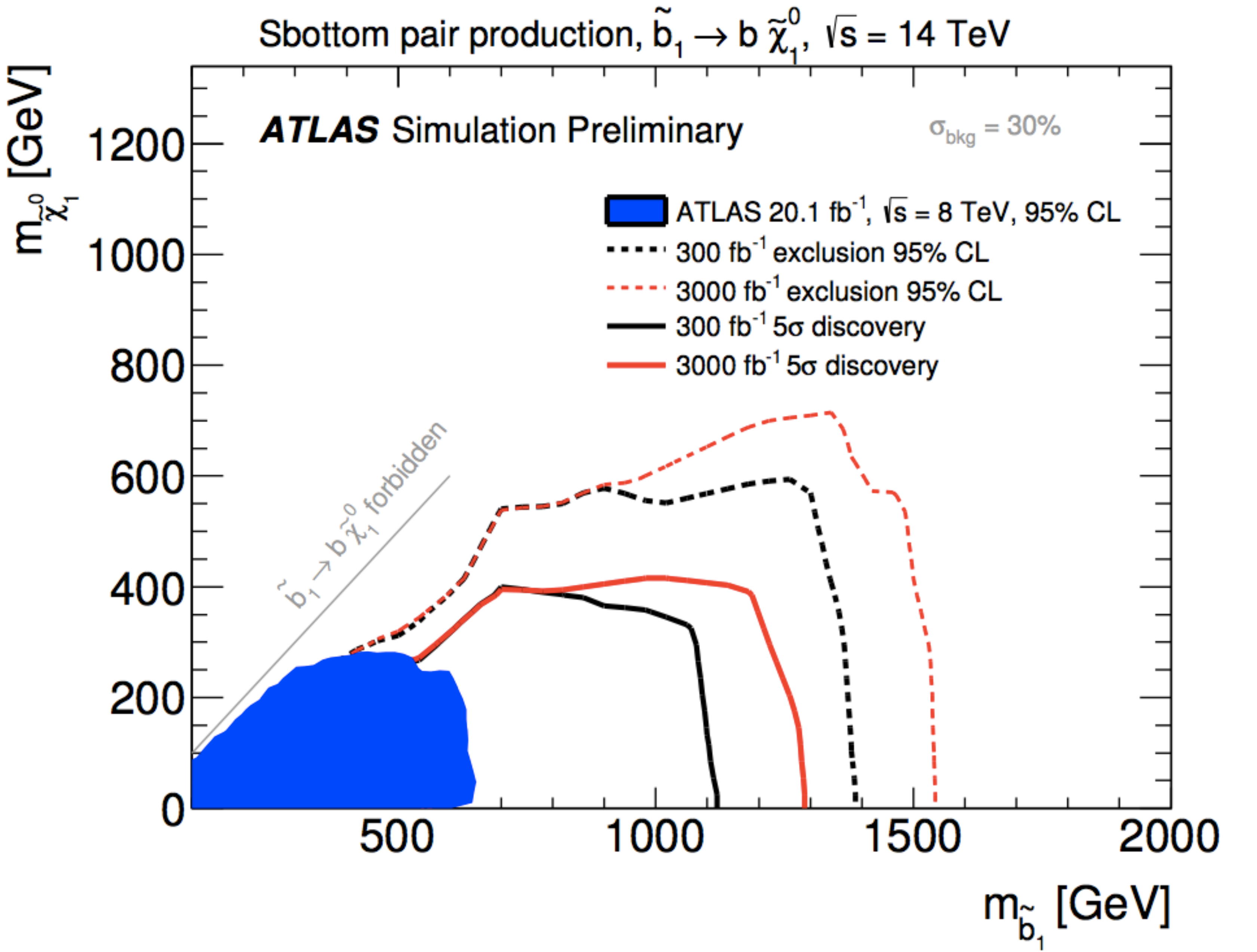}
   \caption{\small Projected evolution with luminosity of the exclusion and
     discovery reach for bottom squarks at the LHC~\cite{ATLAS-SUSY}.}
   \label{fig:sbottom}
\end{figure}

Another example is given in Fig.~\ref{fig:stop}, showing the
luminosity evolution
of the discovery reach at 100~\tev\ for top squarks. The upper mass
reach goes from 6 to 8~\tev\ for $L=3\to30$~\iab, consistent with the
statistical scaling shown in Fig.~\ref{fig:QQreach}. The coverage in
the rest of the $(m_{\tilde{t}} , m_{\tilde{\chi}^0})$ plane does not grow as
rapidly. It might be improved by further optimization of the analyses, and improvements in detector-performance. 

\begin{figure}[htb]
   \centering
\epsfig{width=0.32\textwidth,figure=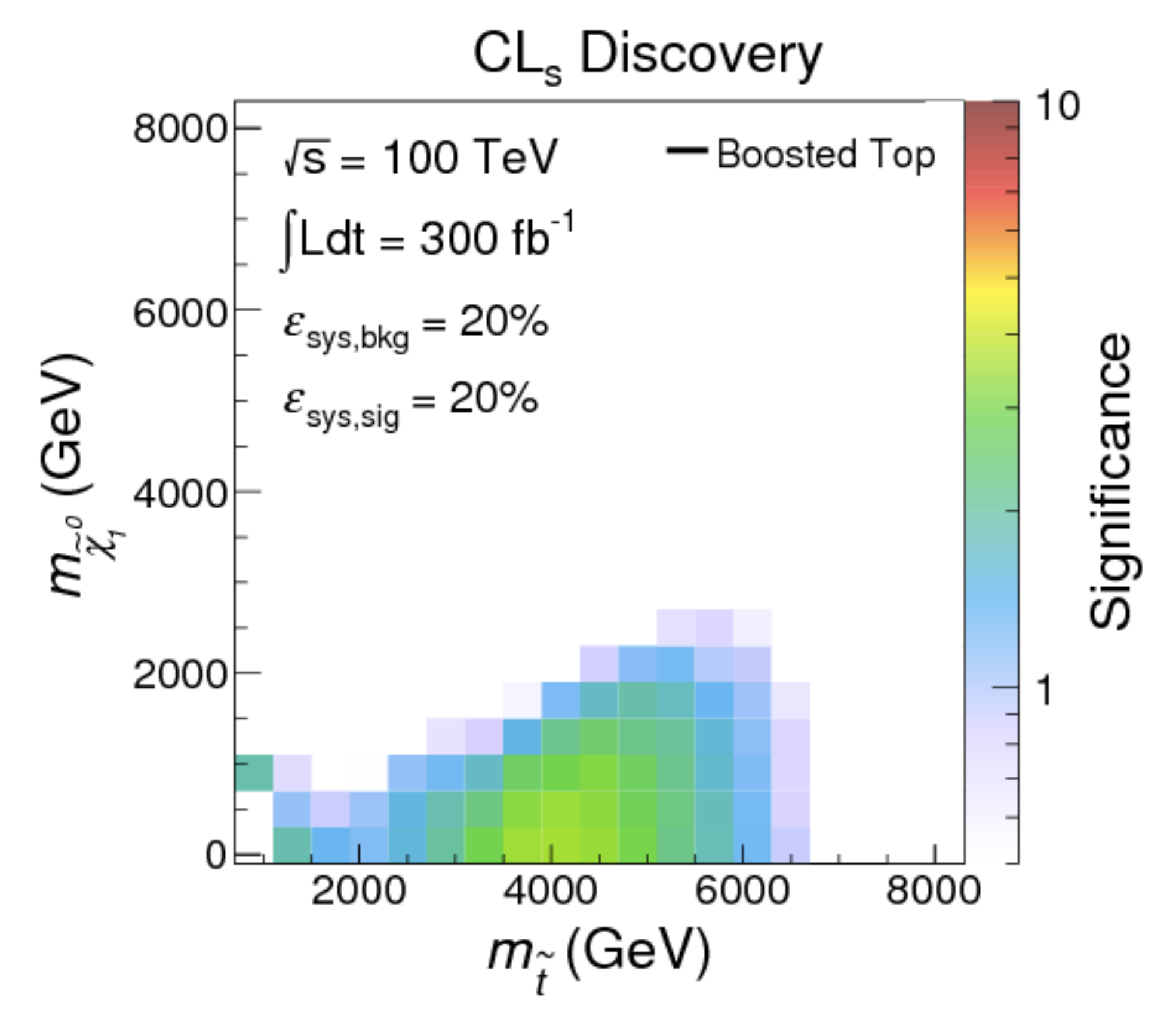}
\epsfig{width=0.32\textwidth,figure=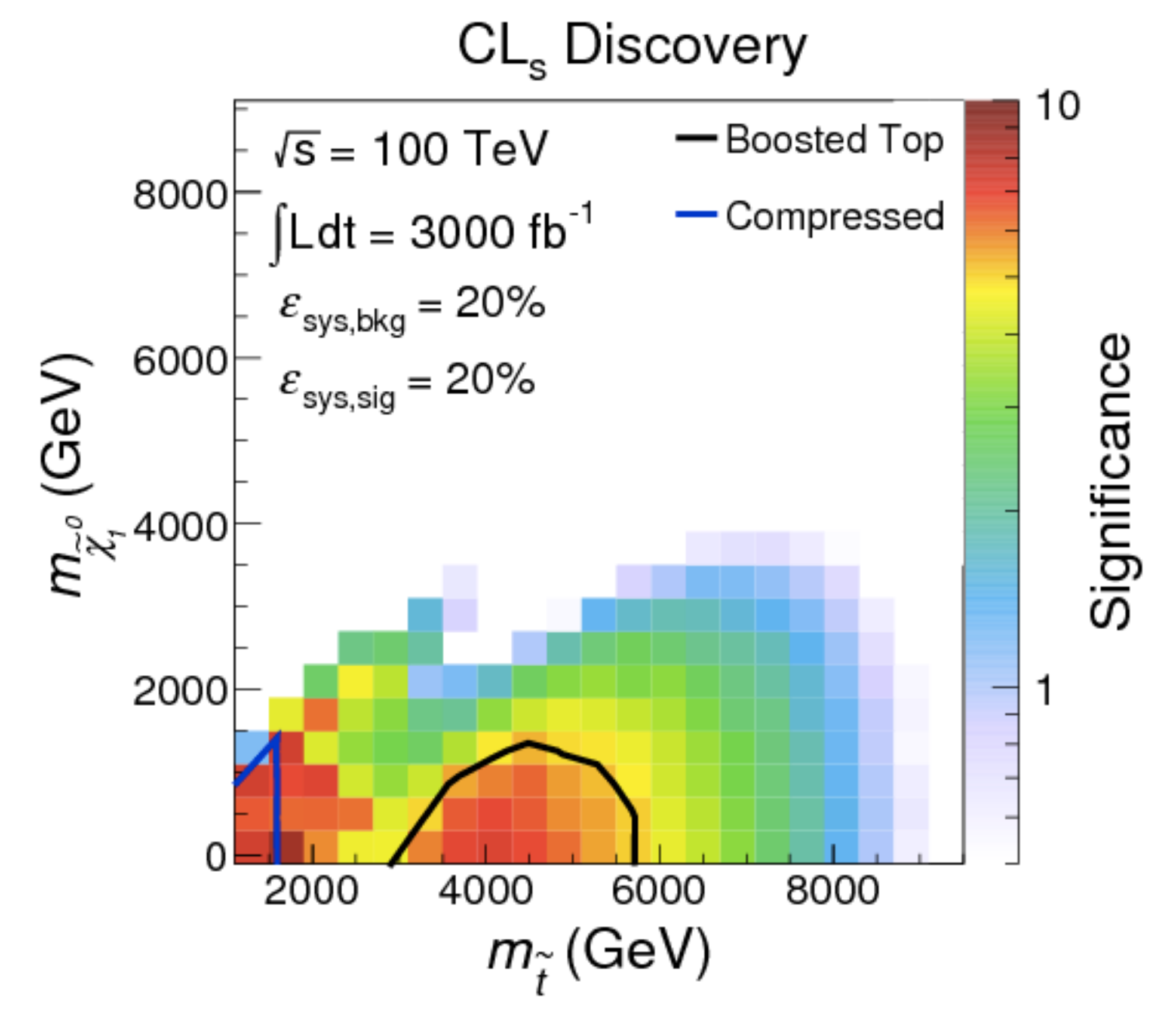}
\epsfig{width=0.32\textwidth,figure=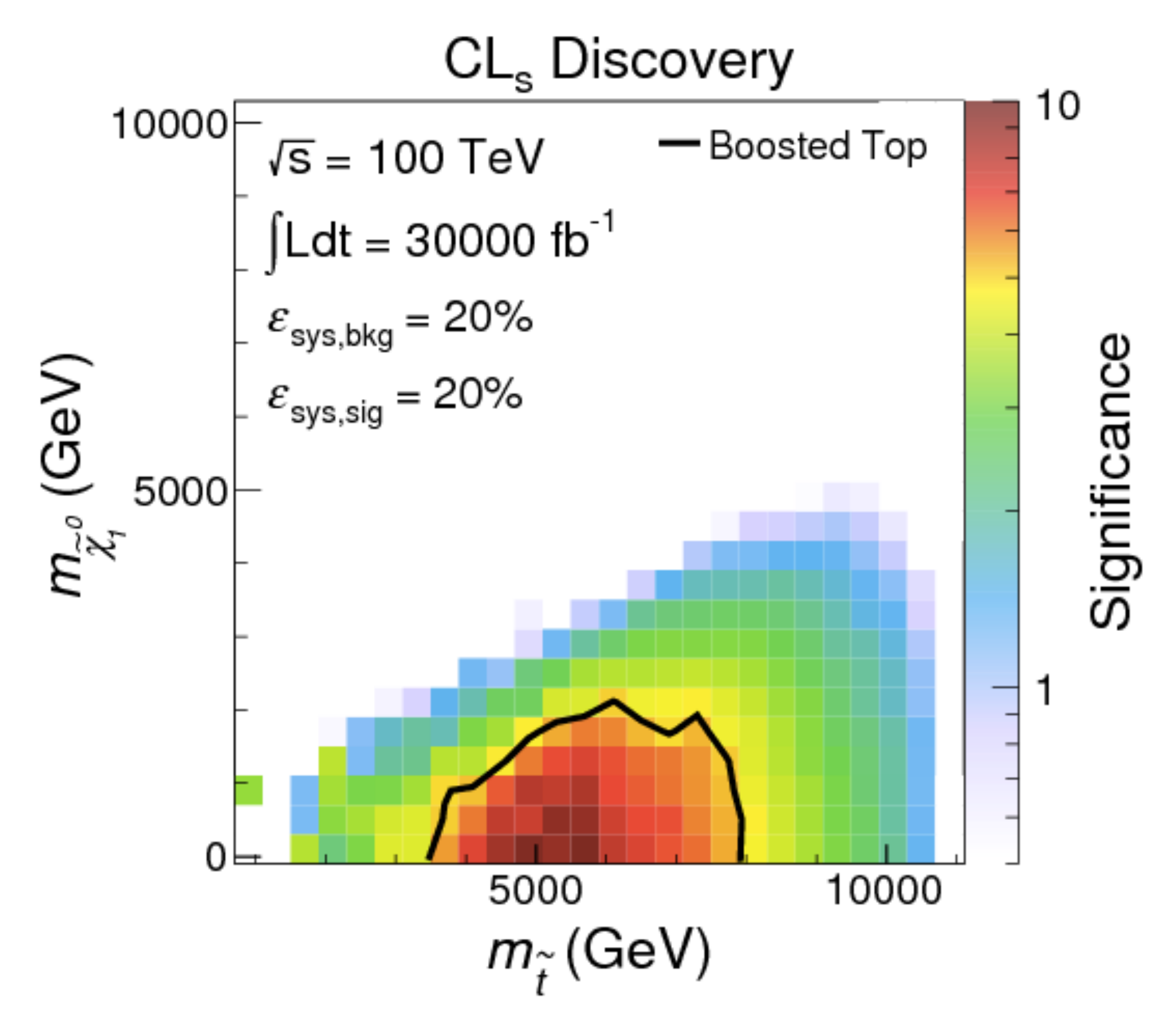}
   \caption{\small Top squark signal efficiency at 100 TeV, with 0.3,
     3 and 30~\iab\ (left to right, respectively), from
     Ref.~\cite{Cohen:2014hxa}}
   \label{fig:stop}
\end{figure}

These examples show that, for the exploration of physics at mass
scales well below the kinematic limit, no generic scaling argument for luminosity can be given.  In particular, for mass scales that
are accessible to the LHC, one should recall that
the increase in energy to 100 TeV will by itself lead to a substantial  increase
in production rates.

\subsection{Precision studies of particles accessible to the LHC}
If the LHC discovers new particles during its future runs, the
production rates may not be sufficient to provide adequate
precision in the determination of their properties. The 100-TeV
collider should then aim to become a ``factory'' environment for these
studies. 
\begin{figure}[htb]
   \centering
\epsfig{width=0.5\textwidth,figure=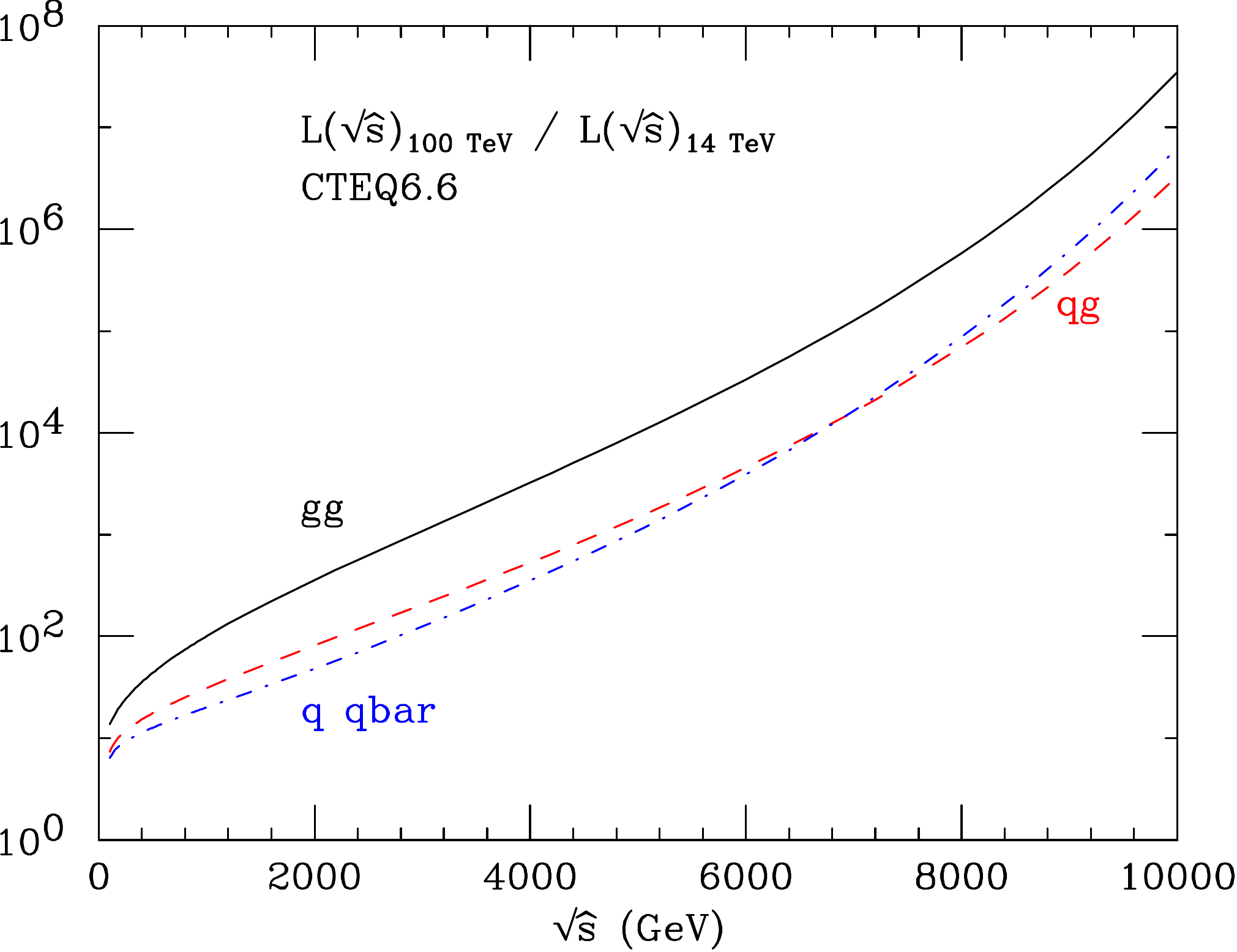}
   \caption{\small Ratio of partonic luminosities at 100 and 14
     TeV, as a function of partonic center-of-mass energy $\sqrt{\hat{s}}$, 
     for different partonic initial states. PDF set
     CTEQ6.6~\cite{Nadolsky:2008zw}, $Q^2=\hat{s}$.
}
   \label{fig:PDF}
\end{figure}

Consider, for example, particles at the upper limit of the HL-LHC
discovery range, for example a gauge boson of mass around parton
subenergy $\sqrt{\hat{s}}=6~\tev$ produced singly in the $q\bar{q}$
channel, or pair production of $\sim$~3~\tev\ particles in the $gg$
channel. Figure~\ref{fig:PDF} shows the partonic luminosity ratios for
various initial-state production channels ($gg$, $q\bar{q}$, $qg$). In
particular, in the cases at hand of $q\bar{q}$ and $gg$ we obtain a
cross-section increase of $10^4$ and $10^5$, respectively. When
accompanied by an increase in integrated luminosity by a factor of
$\sim 10$, this implies event samples up to a million times larger.

In the case of lighter particles, e.g., 1~\tev\ for a resonance in the
$q\bar{q}$ channel or 500~GeV for pair production in the $gg$ channel,
the rate increase due to the partonic luminosities is a factor of
approximately 100. Once again, at low values of $\sqrt{\hat{s}/s}$, an
increase in luminosity by an order of magnitude may be as
advantageous as an increase in energy by a factor of seven. At high
values of $\sqrt{\hat{s}}$ there is a decisive advantage to increasing
$\sqrt{s}$.

\subsection{Study of Higgs-boson properties}
The Higgs-boson inclusive production rate increases from 14 to 100 TeV
by a factor in the range of 10--60, depending on the specific
production process (see Table~\ref{tab:H}). These factors, together
with the improvements in the theoretical systematics and the detector
performance that one can confidently anticipate over the next 30
years, are large enough to promise an important improvement in the
precision with which the Higgs properties can be studied at 100 TeV,
even with a luminosity comparable to that of the LHC. It will be
particularly true of channels such as associated production with top
quarks, $gg\to t\bar{t}H$, and Higgs pair production in gluon fusion,
$gg\to HH$, where the rate increases are the largest (60 and 40,
respectively).

\begin{table}[t]
\centering
 \begin{tabular}{|l|cccccc|}
 \hline
Process & $gg\to H$ & $q\bar{q}\to WH$ & $q\bar{q}\to WH$  & $qq\to qqH$  &
 $gg/q\bar{q} \to t\bar{t}H$ & $gg\to HH$ \\
 \hline
$\sigma(100~\tev)/\sigma(14~\tev)$ &
14.7 & 9.7 & 12.5 & 18.6 & 61 & 42 \\
 \hline
 \end{tabular}
\caption{\small 
Ratio of cross sections at $\sqrt{s}=100$~\tev\ relative to
$\sqrt{s}=14$~\tev\ for various Higgs production processes~\cite{HXSWG}.
\label{tab:H} }
\end{table}
In the case of single Higgs production, detailed studies of the actual
precision reach are  lacking, and it is not possible at
this time to anticipate the luminosity values at which systematic
uncertainties will start to dominate.  Preliminary
studies~\cite{Yao:2013ika,Barr:2014sga,Azatov:2015oxa} are however
available for $HH$ pair production, which will still be very poorly
probed after completion of the HL-LHC program. A prime goal of $HH$ studies is to extract the Higgs-boson self-coupling with a precision of 5\% of the standard-model expectation.  The preliminary studies suggest that this goal can be
reached with 30~\iab, through the measurement of the cross
section  for Higgs pairs in the channel $HH\to
b\bar{b}\gamma\gamma$.

\section{Minimum goals for luminosity}
Experience shows that no collider ever starts at the ultimate
luminosity. It is interesting, therefore, to evaluate what 
minimum luminosity threshold  opens the door on possible discoveries
at 100 TeV. 
\begin{figure}[htb]
   \centering
\epsfig{width=0.6\textwidth,figure=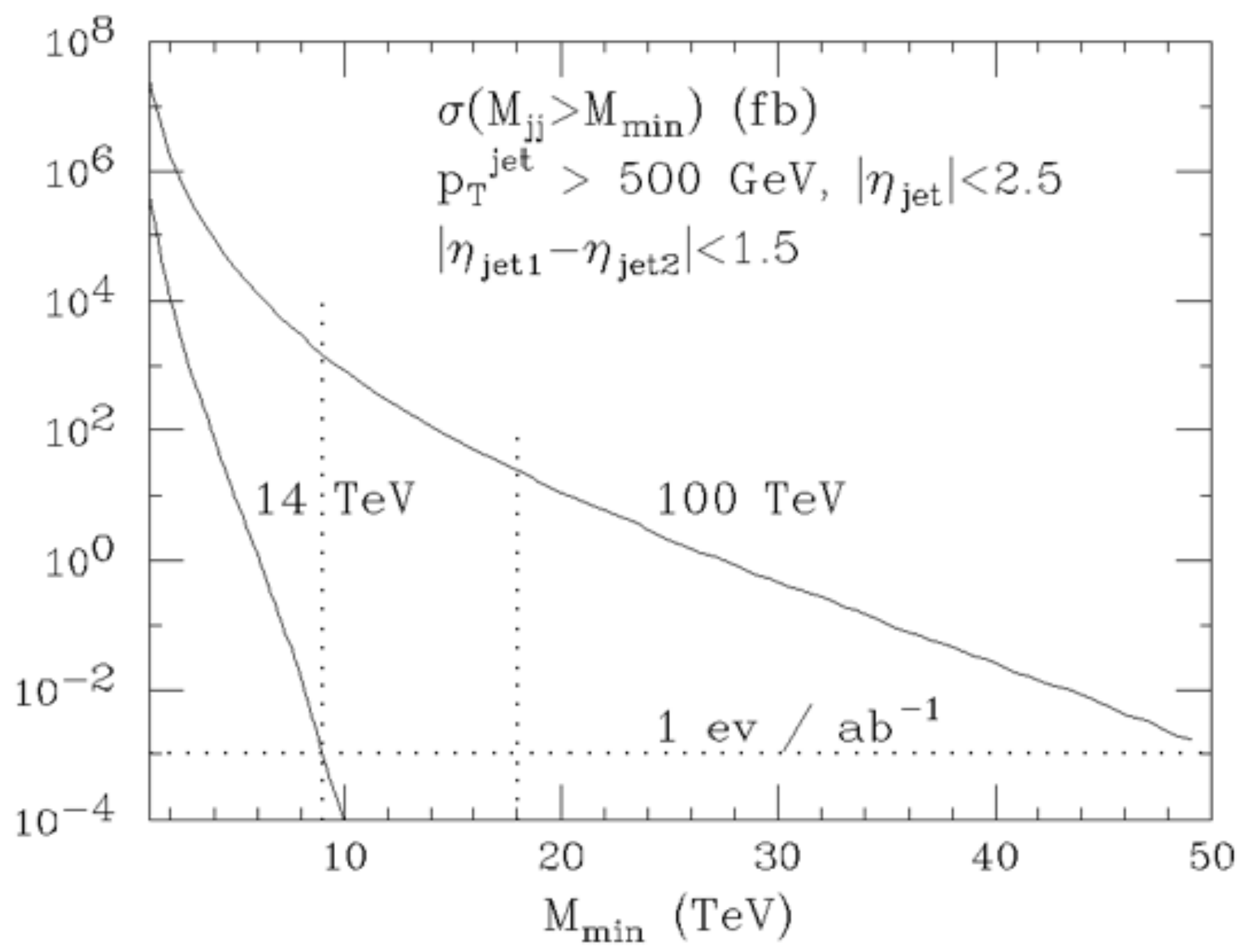}
   \caption{\small Cross sections for the production of dijet pairs
     with invariant mass $M_{jj}>M_{\mathrm{min}}$, at c.m.\ energies $\sqrt{s}=14$ and 100
     TeV. The jets are subject to the $p_T$ and $\eta$ cuts shown in
     the legend.}
   \label{fig:dijets}
\end{figure}

If we consider dijet production as a probe of the shortest distances,
we can extract a reference luminosity target from
Fig.~\ref{fig:dijets}, which shows the leading-order cross section to produce
central dijet pairs as a function of their invariant mass. The
LHC has a sensitivity at the level of 1 event per~\iab\ for dijet
masses above $\sim 9.5$~TeV.  At this mass, the 100 TeV cross section
is 6 orders of magnitude larger, which means that the HL-LHC
sensitivity can be recovered within 1~\ipb, i.e., in
less than a day of running at a luminosity of $10^{32}$~\lum. The
sensitivity to a mass range twice as large, 19~TeV, would require
50~\ipb, namely of the order of one month at $10^{32}$~\lum, and one
year of running at this luminosity would give us events with dijet mass well
above 25~TeV. 

If we consider particles just outside the possible discovery reach of
the HL-LHC, which therefore the LHC could not have discovered, we find
rate increases in the range of $10^4$--$10^5$ that we discussed earlier,
for $q\bar{q}$ and $gg$ production channels, respectively. This means
that integrated luminosities in the range of 0.1--1~\ifb\ are sufficient to push
the discovery reach beyond what the HL-LHC has already explored. This
can be obtained with initial luminosities as small as  $2\times
10^{32}$~\lum.

Finally, we project in Fig.~\ref{fig:massreach_profile} the temporal
evolution of the expansion of discovery reach for various
luminosity scenarios, relative to the reach of 3~\iab\ at 14 TeV. The
left (right) plot shows results for a resonance whose couplings allow
discovery at HL-LHC up to 6~TeV (1~TeV). Once again, we notice that
the benefit of luminosity is more prominent at low mass than at
high mass. We also notice that, considering the multi-year span
of the programme, and assuming a progressive increase of the
luminosity integrated in a year, an early start at low luminosity does
not impact significantly the ultimate reach after a fixed number of
years.
\begin{figure}[htb]
   \centering
\epsfig{width=0.45\textwidth,figure=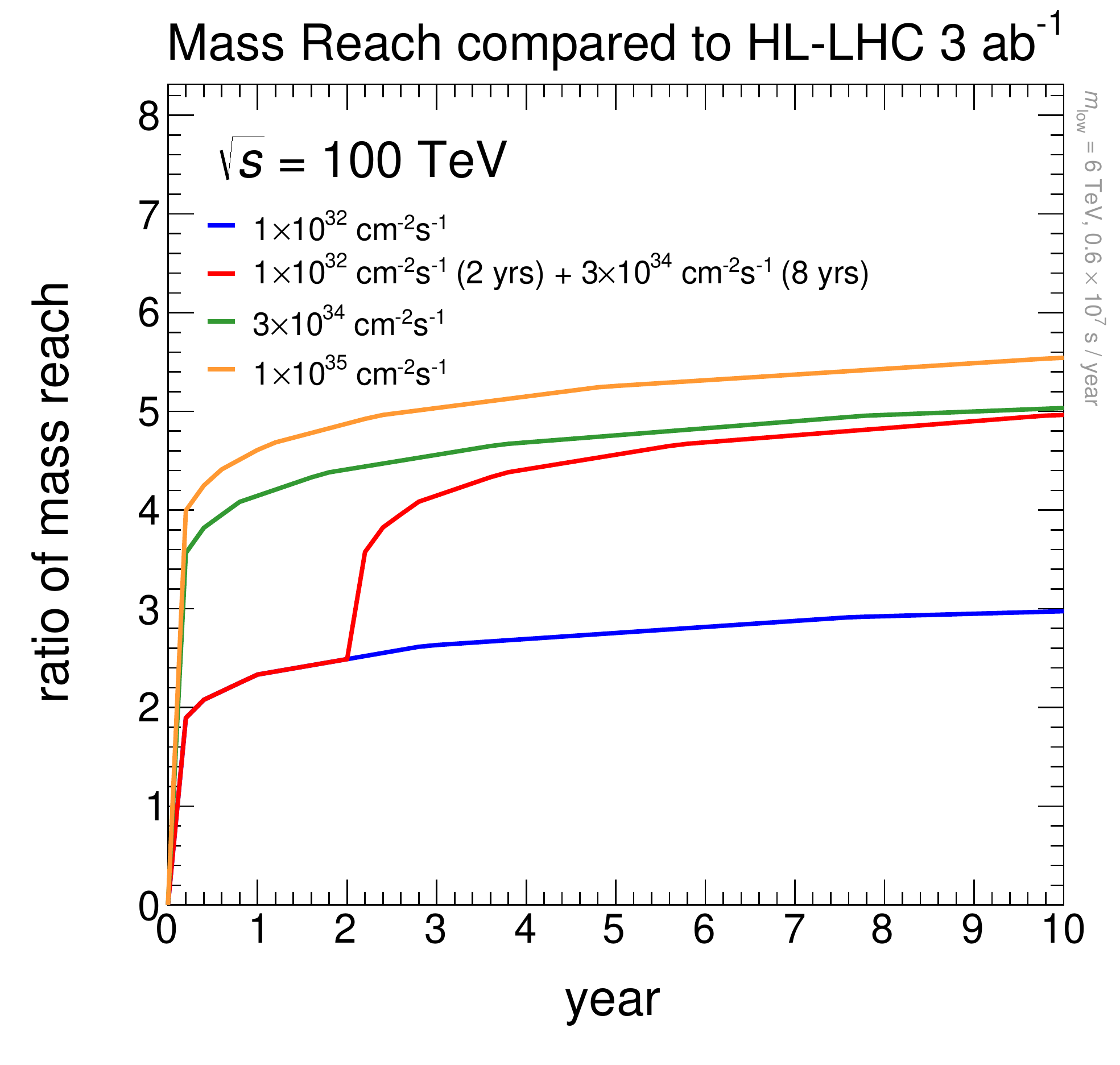}
\hfill
\epsfig{width=0.45\textwidth,figure=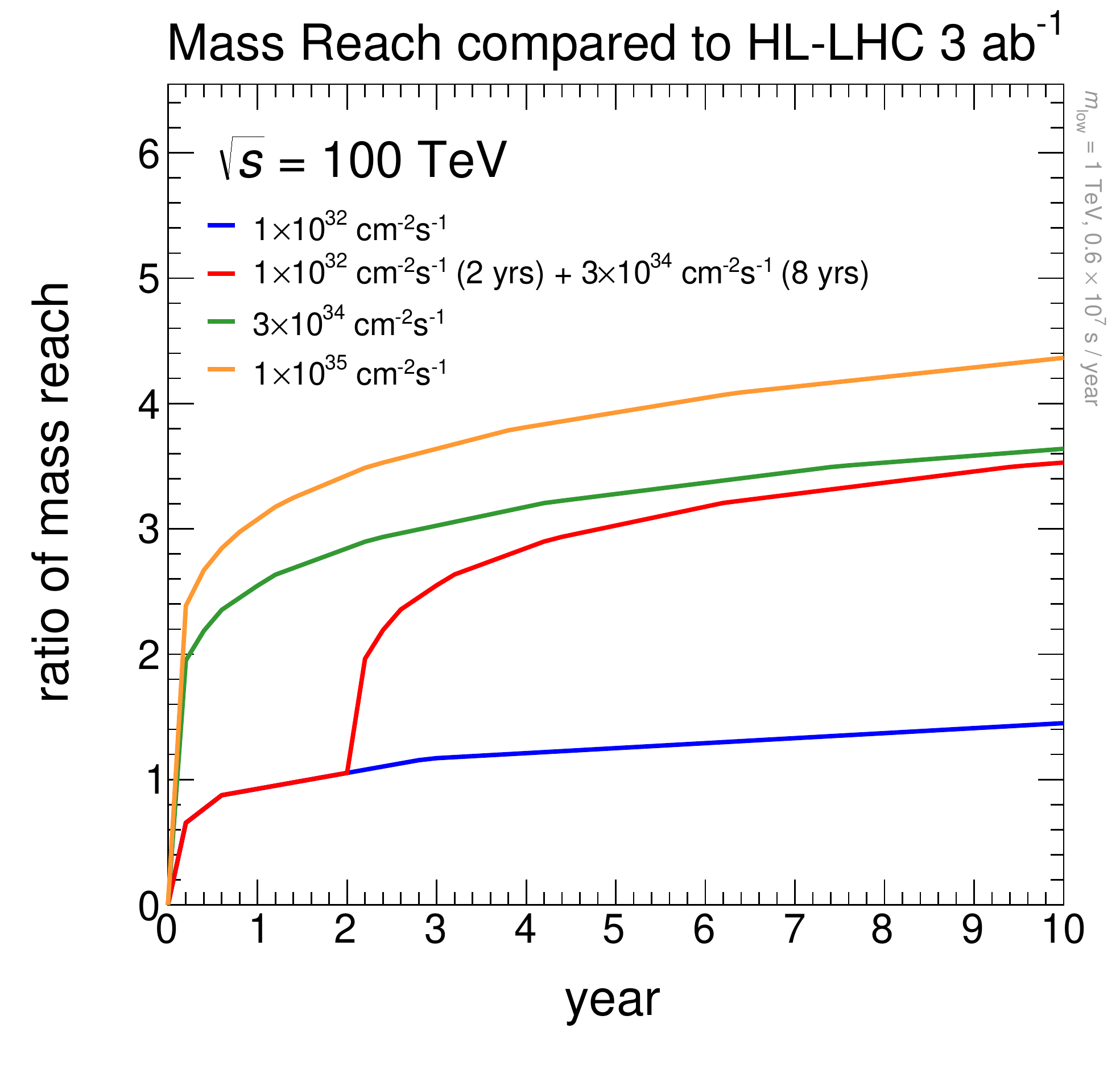}
   \caption{\small Evolution with time of the  mass reach at $\sqrt{s}=100~\tev$, relative to HL-LHC, under different luminosity scenarios (1
     year$\;= 6\times 10^6$ sec).
   The left (right) plot shows the mass increase for a ($q\bar{q}$)
   resonance with couplings enabling HL-LHC discovery at 6~TeV
   (1~TeV).}
   \label{fig:massreach_profile}
\end{figure}

These results are not an argument for modest luminosity as an ultimate goal, but a reminder of the advantages of high collider energy. Should specific very-high-mass targets arise, the overall optimization of energy and luminosity need not be restricted to a single parameter.

\section{Recommendations}
The goal of an integrated luminosity in the range of 10-20~\iab\ per
experiment, corresponding to an ultimate instantaneous luminosity
approaching $2\times 10^{35}$~\lum~\cite{benedikt}, seems well-matched
to our current perspective on extending the discovery reach for new
phenomena at high mass scales, high-statistics studies of possible new
physics to be discovered at (HL)-LHC, and incisive studies of the
Higgs boson's properties.  Specific measurements may set more
aggressive luminosity goals, but we have not found generic arguments
to justify them. The needs of precision physics arising from new
physics scenarios to be discovered at the HL-LHC, to be suggested by
anomalies observed during the $e^+e^-$ phase of a future circular
collider, or to be discovered at 100~TeV, may well drive the need for
even higher statistics. Such requirements will need to be established
on a case-by-case basis, and no general scaling law gives a robust
extrapolation from 14 TeV. Further work on \textit{ad hoc} scenarios,
particularly for low-mass phenomena and elusive signatures, is
therefore desirable.

For a large class of new-physics scenarios that may arise from the
LHC, less aggressive luminosity goals are  acceptable as
a compromise between physics return and technical or experimental
challenges. In particular, even luminosities in the range of
$10^{32}$~\lum\ are enough to greatly extend the discovery reach of
the 100~TeV collider over that of the HL-LHC, or to enhance the precision in the measurement
of discoveries made at the HL-LHC.


\bigskip

{\bf\noindent Acknowledgments \\} This document grew out of
discussions held at the Jockey Club Institute for Advanced Study of
the Hong Kong University of Science and Technology, during the
Programme on \textit{The future of high energy physics,} January 5-30,
2015. We thank Henry Tye and members of the Institute for the
hospitality, the participants for contributing to a stimulating
environment, and Prudence Wong for helpful practical assistance.  In
particular, we acknowledge informative discussions with Stephen
Gourlay, Ian Low, Vladimir Shiltsev, Dick Talman, Weiming Yao and
Charlie Young, and continuous encouragement from Michael Benedikt and
Weiren Chou.  The work of MLM was performed in the framework of the
ERC grant 291377, ``LHCtheory: Theoretical predictions and analyses of
LHC physics: advancing the precision frontier''. Fermilab is operated
by Fermi Research Alliance, LLC, under Contract No.  DE-AC02-07CH11359
with the United States Department of Energy.  CQ thanks John
Iliopoulos and the \textit{Fondation Meyer pour le d\'eveloppement
  culturel et artistique} for generous hospitality. The work of IH was
supported in part by the Office of Science, Office of High Energy Physics,
of the U.S. Department of Energy under contract DE-AC02-05CH11231.

\appendix
\section{Scaling relations}
The cross section $\sigma$ is 
\begin{eqnarray}
\sigma &\sim& L_{\rm p} \cdot \hat{\sigma} \nonumber \\
&\sim& \frac{1}{\tau^a} \hat{\sigma}, 
\end{eqnarray}
where $\hat\sigma$ is the partonic cross section, and we have assumed
that the parton luminosity $L_{\rm p}$ falls as a power law with
increasing $\tau = \hat{s}/s$. In the signal process where the new
physics particle mass scale is $M$, we will further assume that
\begin{equation}
\hat{\sigma} \propto \frac{1}{M^2}.
\end{equation}
Next, we consider two different colliders with center of mass energies
$\sqrt{s_1}$ and $\sqrt{s_2}$, with integrated $pp$ luminosity
${\cal{L}}_1$ and ${\cal{L}}_2$, respectively. We assume the mass
reaches of new physics at those two colliders are $M_1$ and $M_2$,
respectively. The corresponding parton fractions are $\tau_i = M_{i}^2
/s_{i}$, $(i=1,2)$.  Assuming that the reach is obtained by the same number of signal events, we have
\begin{eqnarray}
\frac{1}{\tau_1^{a}} \frac{1}{M_1^2} {\cal{L}}_1 = \frac{1}{\tau_2^{a}} \frac{1}{M_2^2} {\cal{L}}_2,
\end{eqnarray}
which means 
\begin{equation}
\frac{M_2}{M_1} = \left( \frac{s_2}{s_1} \right)^{\frac{a}{2 a +2}} \left( \frac{{\cal{L}}_2}{{\cal{L}}_1} \right)^{\frac{1}{2 a +2}} .
\end{equation}
For large $a$, this means energy is more important, and the gain
with luminosity can be quite slow. In particular, if we require
$M_2/M_1 = E_2 / E_1$, we need ${\cal{L}}_2 = (E_2/E_1)^2
{\cal{L}}_1$, as emphasized in
Refs.~\cite{Barletta:2014vea,Richter:2014pga}. 
However, this slow gain with luminosity also means that one would not lose too much mass reach by going to a much lower luminosity. As demonstrated here, this is ultimately due to the fact that the parton luminosity is steeply falling, in particular near the edge of the kinematical reach of a collider.  The gain with luminosity is more important for smaller $\alpha$ or lower $\tau$ (lower mass).

\begin{figure}[h!]
\begin{center}
\includegraphics[scale=0.5]{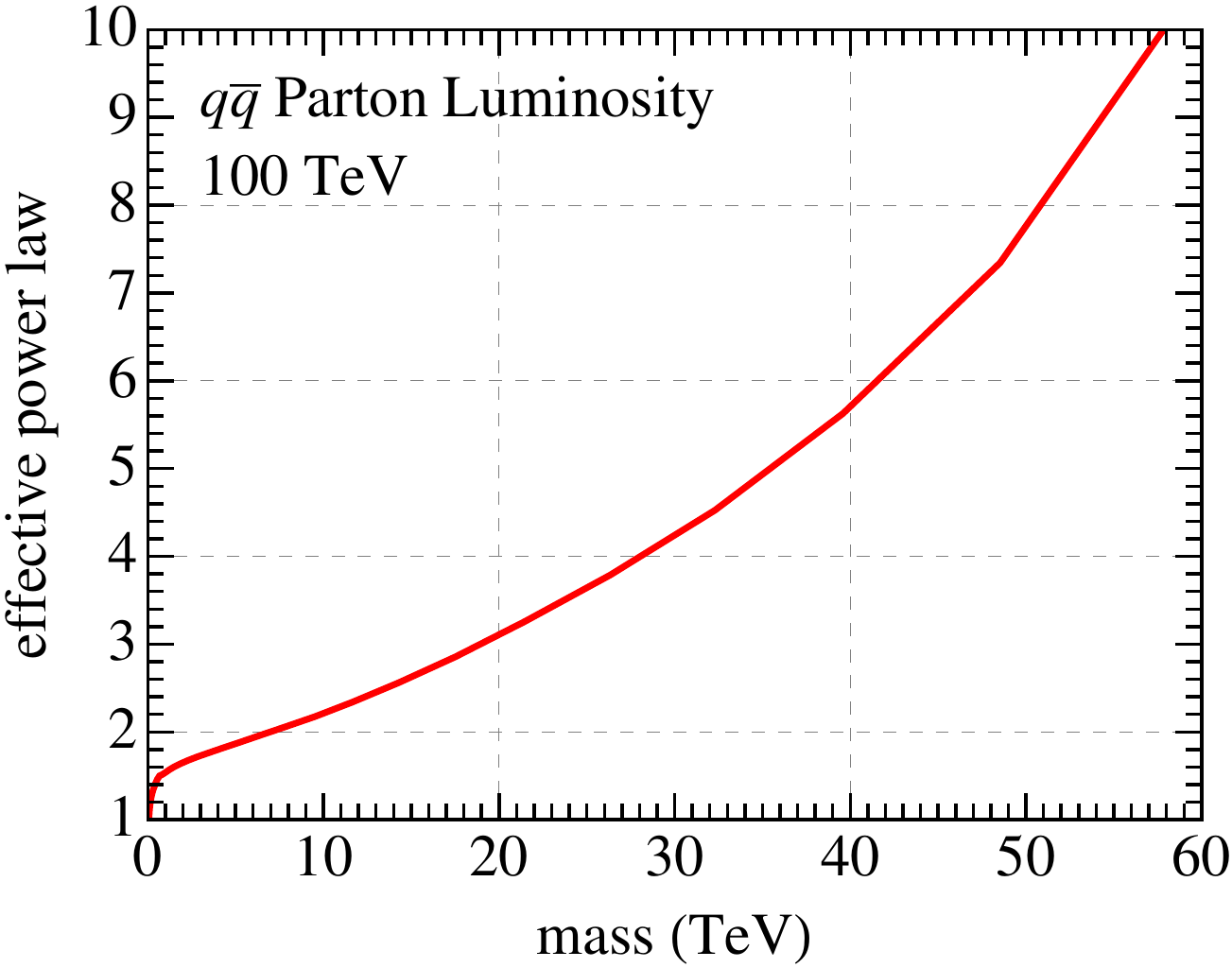} \ \ \includegraphics[scale=0.5]{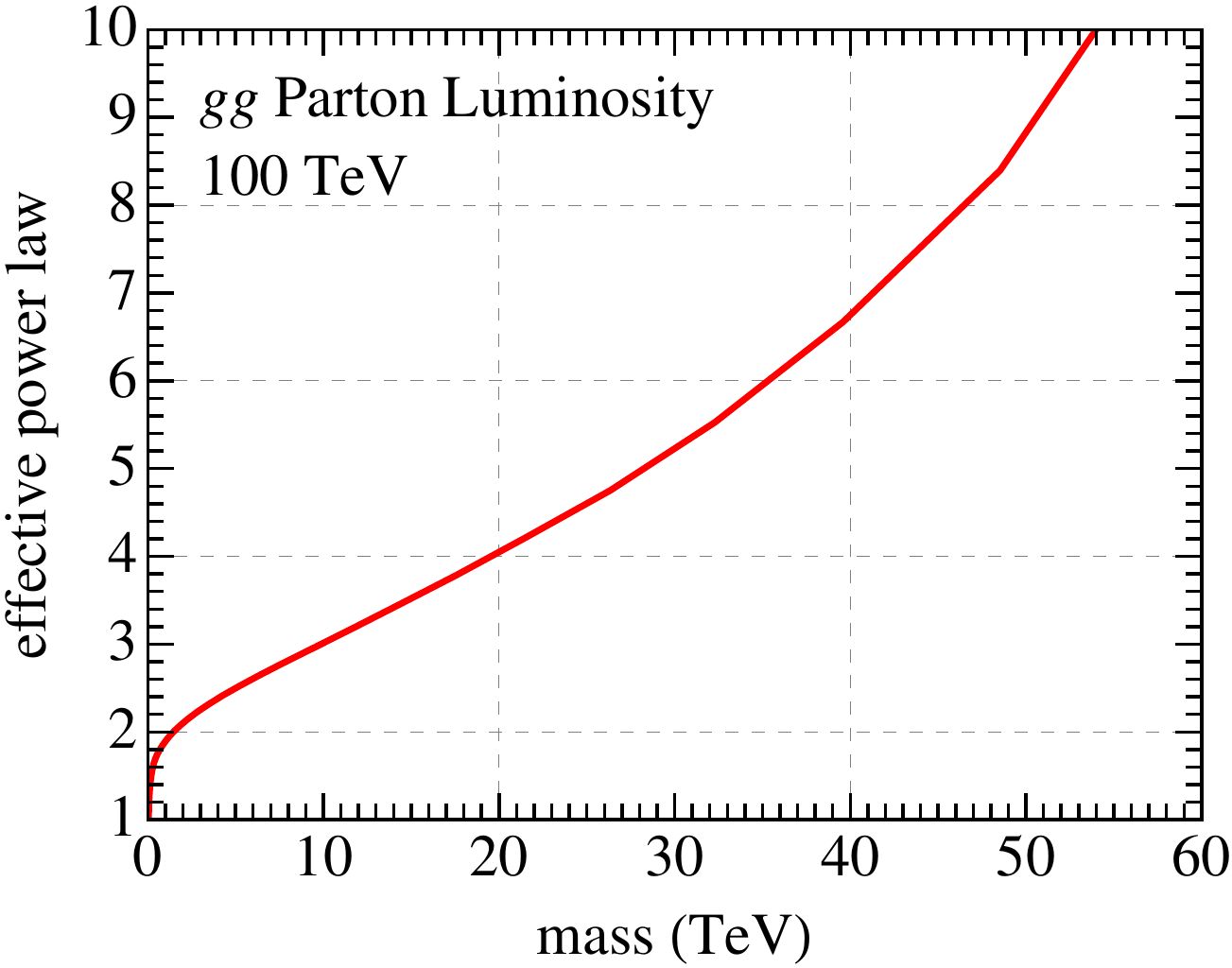}

\end{center}
\caption{\label{fig:atau}The dependence of power $a$ on mass scale $M = \sqrt{\hat{s}} = \sqrt{s\tau} $}
\end{figure}

Some obvious approximations are made here. First of all, we ignored
anomalous scaling. We also assumed that for the relevant range of
$\tau$, $a$ remains approximately constant. This is certainly not true
for full range of $\tau$. However, $a$ does not vary too steeply with
$\tau$, see Fig.~\ref{fig:atau}. For comparing reaches, we often
consider similar values of $\tau$. 

Next we consider the gain luminosity with the same collider, i.e., $E_1 = E_2$. We have
\begin{equation}
\frac{M_2}{M_1} = \exp \left(\frac{1}{2a +2}  \log({\cal{L}}_2 /{\cal{L}}_1)\right) \simeq 1 + \frac{1}{2a +2}  \log({\cal{L}}_2 /{\cal{L}}_1),
\end{equation}
or
\begin{equation}
M_2 -M_1  \simeq \frac{M_1}{2 a+2} \log({\cal{L}}_2 /{\cal{L}}_1)
\end{equation}
For example, considering $q \bar{q}$ initial state, around $M_1 \simeq 40 $ TeV, $a \simeq 5.5$ (from Fig.~\ref{fig:atau}), we have approximately
\begin{equation}
M_2 - M_1 \sim (7 \mbox{ TeV}) \times \log_{10}  ({\cal{L}}_2 /{\cal{L}}_1) 
\end{equation}
At the same time, for lower mass $M_1 \simeq 20 $ TeV, $a \simeq 3$, we have instead
\begin{equation}
M_2 - M_1 \sim (5.5 \mbox{ TeV}) \times \log_{10} ({\cal{L}}_2 /{\cal{L}}_1) 
\end{equation}


\end{document}